
\def\tr{{\rm tr}}
\def\ca{{\cal A}}
\def\br{{\bf R}}
\def\bz{{\bf Z}}
\def\cg{{\cal G}}
\def\aut{{\cal A}\!ut}
\def\lieg{{\bf g}}
\def\sigm{\Sigma M}
\def\sigp{\Sigma P}
\def\sigg{\Sigma G}
\def\sigph{\widehat{\Sigma P}}
\def\sigpt{\widetilde{\Sigma P}}
\def\siggh{\widehat{\Sigma G}}

\font\titlefont=cmb10 scaled\magstep1
\magnification=\magstep1
\baselineskip=15pt
\line{\hfil SISSA 36/93/EP}
\line{\hfil hep-th/9304054}
\line{\hfil March 1993}
\vskip  2.4truecm
\centerline{\titlefont GLOBAL ASPECTS OF p--BRANES}
\vskip  1truecm
\centerline{\phantom{$^1$,$^2$}\bf J. Mickelsson
\footnote{$^1$}{\it Permanent address: Department of
Mathematics, University of Jyv\"askyl\"a\hfil\break
SF-40100, Jyv\"askyl\"a, Finland},
\footnote{$^2$}{\it Supported in part by Alexander von Humboldt Stiftung}}
\medskip
\centerline{\it Institut f\"ur Theoretische Physik, Universit\"at Freiburg}
\centerline{\it Hermann--Herder--Stra\ss e 3, D-7800, Freiburg}
\bigskip
\centerline{\bf R. Percacci}
\medskip
\centerline{\it International School for Advanced Studies,
via Beirut 4, 34014 Trieste, Italy}
\centerline{\it and Istituto Nazionale di Fisica Nucleare,
Sezione di Trieste.}
\bigskip
\vskip  1.4truecm
\centerline{\bf Abstract}
\smallskip
\midinsert\narrower{
We generalize to dimension $p>1$ the notion of string structure and
discuss the related obstruction.
We apply our results to a model of bosonic $p$-branes propagating on a
principal $G$-bundle, coupled to a Yang--Mills field and an
antisymmetric tensor field and in the presence of a Wess-Zumino
term in the Lagrangian.
We construct the quantization line bundle and discuss the action of
background gauge transformations on wave functions.}
\endinsert
\vfill\eject

\beginsection{1. INTRODUCTION}

The notion of string structure is an elegant way of formulating the
absence of certain anomalies for superstrings coupled to
gauge fields in the target space [1,2].
In this paper we will discuss the generalization of this notion to
$p$-branes with arbitrary odd $p$.

Let us begin by choosing a $p$-dimensional manifold $\Sigma$ which
is compact, connected, oriented and without boundary, and a
``target space'' $M$, to be interpreted as spacetime.
A $p$-brane is a map from $\Sigma$ to $M$.
In order to describe the coupling of the $p$-brane to a Yang--Mills field,
we choose a principal bundle $P$ with compact structure group $G$
and base space $M$, with connection defined by some connection form
$\alpha$. We adopt the standard convention
that $G$ acts freely on $P$ from the right.
Let $\sigm$ be a shorthand for $Maps(\Sigma,M)$, and similarly with
$P$ and $G$ (`$Maps$' will always mean the space of smooth maps between
the given domain and target). The configuration space of the
``$p$-branes with internal symmetry '', $\sigp$, is
a principal bundle over $\sigm$ with structure group $\sigg$.
The right action of $g\in\sigg$ on $\varphi\in\sigp$ is given by the
pointwise action of $G$ on $P$:
$$
(\varphi g)(\sigma)=\varphi(\sigma)g(\sigma)\ .\eqno(1.1)
$$
In the case of the string, the loop group $S^1 G$ has  central
extensions $\widehat{S^1 G}$. A string structure is a prolongation
of the principal $S^1 G$ bundle $S^1 P$ over $S^1 M$ to a principal
$\widehat{S^1 G}$ bundle $\widehat{S^1 P}$.
Note that $\widehat{S^1 P}$ is also a circle bundle over $S^1 P$.

In the case $p>1$, $p$ odd, the group $\sigg$ also has extensions,
but unlike the case of the string the interesting extensions are not simply
central
extensions. In the setting described above, the natural extensions
are parametrized by connections in $P$ and have as fiber the space
of $S^1$-valued functionals on $\sigp$. We will denote
$\siggh_\alpha$ the extension defined by the connection $\alpha$.
(These extensions will be described more precisely in sections 2-4.)
A priori there seem to be two natural ways of generalizing the notion of
string structure to higher $p$-branes. One could define a ``$p$-brane
structure'' as a prolongation of $\sigp$ to a principal bundle
$\sigph$ over $\sigm$ with structure group $\siggh_\alpha$;
or one may define it as a principal $S^1$ bundle $\sigpt$ over $\sigp$,
together with a faithful right action of $\siggh_\alpha$ on $\sigpt$.

Both notions agree in the case $p=1$. We shall follow here the second
approach. Note that $\sigpt$, as a fiber
bundle over $\sigm$, has a standard fiber isomorphic to a circle bundle
over $\sigg$ and is not a principal bundle.

The obstruction to our construction of a ``$p$-brane
structure'' is given by
a certain characteristic class of the bundle $P,$ which depends on the
class of the extension $\siggh_{\alpha}.$
For the extensions that we shall consider, the obstruction is
$$
c=k_p\tr\,\phi^{p+3\over2}\ ,\eqno(1.2)
$$
where $\phi=d\alpha+{1\over2}[\alpha,\alpha]$
is the curvature form of $\alpha$, $\tr$ is the trace in a fixed
finite dimensional representation of $G$ and $k_p$ is a
normalization factor chosen in such a way that the integral of $c$
over any $(p+3)$-dimensional compact manifold
without boundary is an integer. (If $G=SU(N)$, $c$ is the
Chern class $c_{p+3\over2}$). Other polynomial invariants can also be
used, leading to different group extensions and prolongations.

In the case of the string, the existence of a string structure can be
viewed as a condition for the existence of the Dirac--Ramond operator [1].
This operator is related to the supersymmetry charge.
It is reasonable to expect that also in the case $p>1$ the existence
of a ``$p$-brane structure'' is a condition for the existence of an
analogue of the Dirac--Ramond operator.
This operator presumably arises in a supersymmetric theory of
$p$-branes in which the degrees of freedom in the fibers of $P$ are
replaced by suitable chiral fermions.
However, in the absence of a supersymmetric Lagrangian for $p$-branes
coupled to Yang--Mills, we shall not discuss these aspects here.
Instead, we shall consider a particular model dynamics for bosonic
$p$-branes which was recently discussed [3].
At the local level it involves, in addition to
the Yang--Mills field $A$, also a $(p+1)$-form $B$.
Invariance under target space gauge transformations (automorphisms
of $P$) results in this model from a kind of Green--Schwarz anomaly
cancellation mechanism in which the form $B$ plays a crucial role.
Unlike the case $p=1$, this bosonic theory is not expected to be
equivalent to its fermionic counterpart, but it reproduces faithfully
features related to the cancellation of anomalies.
We find that the construction of the invariant action
requires that the characteristic class $c$ vanishes.
This is the manifestation at the Lagrangian level of the
obstruction discussed above.

Starting from the invariant action, we construct the circle bundle
$\sigpt_{(\alpha,H)}$ with right $\siggh_\alpha$ action (the construction
depends on one additional datum, namely the
$(p+2)$-form $H$ given in (5.1) below).
This bundle is the quantization bundle; the
sections of the associated complex line bundle are the Schr\"odinger
wave functions.
We then show that the group of target space gauge transformations
(automorphisms of $P$) can be realized on wave functions without
extension.
Our findings agree with the results of explicit canonical calculations
in the case of a trivial bundle $P=M\times G$ [4,5].

\beginsection{2. EXTENSIONS OF THE LIE ALGEBRA $\Sigma\lieg$}

Let $\lieg$ be the Lie algebra of $G$ and $\Sigma\lieg=
Maps(\Sigma,\lieg)$ the Lie algebra of $\sigg$, under pointwise commutators.
Also, denote $\ca$ the space of $\lieg$-valued one-forms on $\Sigma$.
There exist nontrivial extensions of $\Sigma\lieg$ by the
abelian ideal $Maps({\cal A},i\br)$ [6,7].
These extensions can be described as follows.
As a vector space, the extended Lie algebra is the direct sum
$\Sigma\lieg\oplus Maps(\ca,i\br)$. The Lie bracket is then given by
$$
[(X,\gamma),(X',\gamma')]=
([X,X'],\delta_X\gamma'-\delta_{X'}\gamma
+c_2(.;X,X'))\ .\eqno(2.1)
$$
Here $c_2$ is a two-cocycle in $\Sigma\lieg$ with values in
$Maps(\ca,i\br)$, {\it i.e.} it satisfies:
$$
\delta_X c_2(.;Y,Z)-c_2(.;[X,Y],Z)+ \rm{ cyclic\ permutations }=0\
,\eqno(2.2)
$$
where $\delta_X$ denotes the infinitesimal gauge variation of a
functional of $A$. The cocycle can be written as
$$
c_2(A;X,Y)=\int_\Sigma\omega^2_p(A;X,Y)\ ,\eqno(2.3)
$$
where $\omega^2_p$ is a $p$-form on $\Sigma$ depending polynomially on
the vector potential $A\in\ca$.
The form $\omega^2_p$ can be obtained by the dimensional descent
procedure from an invariant polynomial in the curvature. Starting from
the invariant polynomial given in (1.2), with $\phi$ replaced by
$F=dA+{1\over2}[A,A]$,
one defines the Chern--Simons form $\omega^0_{p+2}(A)$ by
$$
d\omega_{p+2}^0(A)=k_p\tr F^{{p+3}\over2}\ .\eqno(2.4)
$$
The superscript $0$ refers to the degree of $\omega$ as a cochain
on the Lie algebra $\Sigma\lieg$, while the subscript $(p+2)$
refers to its degree as a differential form.
The coboundary of the Chern--Simons form defines (up to an exact form)
the one--cochain $\omega^1_{p+1}$:
$$
\delta_X\omega^0_{p+2}(A)=d\omega^1_{p+1}(A,X)\
.\eqno(2.5)
$$
It can be written
$$
\omega^1_{p+1}(A,X)=\tr\,dX\,\Phi_p(A)\ ,
\eqno(2.6)
$$
where the $p$-form $\Phi_p=\Phi_p^a T_a$ is a polynomial in $A$ and $F$.
For $p=1,3,5$ this polynomial is given by
$$
\eqalignno{
\Phi_1=&-k_1 A\ ,&(2.7a)\cr
\Phi_3=&-{1\over2} k_3(FA+AF-A^3)\ ,&(2.7b)\cr
\Phi_5=&-\!{1\over3}
k_5\!\left[(F^2A+FAF+AF^2)\!-\!{4\over5}(A^3F+FA^3)\!
-\!{2\over5}(A^2FA+AFA^2)\!+\!{3\over5}A^5\right]\! . \quad &(2.7c)\cr}
$$
The coboundary of $\omega^1_{p+1}$ defines $\omega^2_p$:
$$
\delta_{X}\omega^1_{p+1}(A,Y)-
\delta_{Y}\omega^1_{p+1}(A,X)-
\omega^1_{p+1}(A,[X,Y])
=d\omega^2_p(A,X,Y)\  .\eqno(2.8)
$$
In the cases $p=1,3,5$ we have
$$
\eqalignno{
\omega^2_1(A,X,Y)&=
-2k_1\tr\,X dY\, &(2.9a)\cr
\omega^2_3(A,X,Y)&=
 -k_3\tr\, [dX,dY]A\, &(2.9b)\cr
\omega^2_5(A,X,Y)&=
{1\over15}k_5\tr\,
(5F-3A^2)\big(2A[dX, dY]-
dX A dY+dY A dX\big)\ . &(2.9c)\cr}
$$
These are the forms that we shall use in the definition of the
cocycles (2.3).
In particular for $p=1$ this gives the familiar central term of
a Kac-Moody algebra; since it is independent of $A$ in this special
case the fiber can be restricted to the constant functions.

It will be useful to consider also the cocycles $\hat c_2$
differing from $c_2$ by the coboundary of the one-cochain
$\Phi(A,X)=\int_\Sigma \tr X\Phi_p(A)$:
$$
\hat c_2(A,X,Y)=c_2(A,X,Y)-
\left(\delta_X\Phi(A,Y)-\delta_Y\Phi(A,X)-\Phi(A,[X,Y])\right)
\ .\eqno(2.10)
$$
These cocycles can also be written in the form (2.3), with
$$
\eqalignno{
\hat\omega^2_1(A,X,Y)=&\
k_1\tr\,[X,Y]\,A\ ,&(2.11a)\cr
\hat\omega^2_3(A,X,Y)=&\
{1\over2} k_3\tr\,\left([X,Y](FA+AF-A^3)
+X dAY A-X AY dA\right)\ ,&(2.11b)\cr
\hat\omega^2_5(A,X,Y)=&\
{1\over5} k_5\tr\Big\{
[X,Y]\big[2(F^2A+AF^2)+FAF-A^3F-FA^3 \cr
&-A^2FA-AFA^2  \big]  +2(XdAY-YdAX)(FA+AF-A^3)\cr
&-2(XAY-YAX)d(FA+AF-A^3)\Big\}\  .&(2.11c)\cr}
$$
Let us now fix a connection in $P$ with connection form $\alpha$.
The pullback of $\alpha$ by means of the map $\varphi$ is an
element of $\ca$ and the right action (1.1) of $\sigg$ on $\sigp$
induces a gauge transformation on the potential $\varphi^*\alpha$:
$$
(\varphi g)^*\alpha=g^{-1}(\varphi^*\alpha)g+g^{-1}dg\ .\eqno(2.12)
$$
Consider the two-cocycle $c_2'$ with values in $Maps(\sigp,i\br)$
defined by
$$
c_2'(\varphi;X,Y)=c_2(\varphi^*\alpha;X,Y)\ .  \eqno(2.13)
$$
(In the following we will drop the primes for notational simplicity).
It gives rise to an extension of $\Sigma\lieg$ by $Maps(\sigp,i\br)$,
denoted $\widehat{\Sigma\lieg}_\alpha$. These are the extensions
arising in $p$-brane theory.

\beginsection{3. COHOMOLOGY OF $\sigp$}

It will be useful to view the cocycles (2.13) from a
different point of view.
We recall that given an element $X$ of the Lie algebra $\lieg$ of $G$
one can construct a vector field $\tilde X$ on $P$, called a
fundamental vector field, generating the right action of $G$.
This map is an isomorphism from $\lieg$ to the vertical subspace
at each point in $P$. Given a $k$-form
$\beta_k$ on $\sigp$, one can construct a $k$-cochain $c_k$
on $\Sigma\lieg$ with values in $Maps(\sigp,i\br)$ by
$$
c_k(\varphi;X_1,\dots,X_k)=(\beta_k(\varphi))(\tilde X_1,\dots,\tilde X_k)\ .
\eqno(3.1)
$$
Conversely, given a $k$-cochain $c_k$ one can always find a $k$-form
$\beta_k$ on $\sigp$ satisfying (3.1). Clearly this form is not
uniquely defined: only the restriction of $\beta_k$ to the
fibers of $\sigp$ is determined by this condition.
We thus have a surjective map from $k$-forms on $\sigp$ to
$k$-cochains on $\Sigma\lieg$ with values in $Maps(\sigp,i\br)$.
The kernel of this map consists of the forms whose contraction with a
vertical vector vanishes.
Under this map, the exterior differential of $\beta_k$ is mapped to
the coboundary of $c_k$. Closed (resp. exact) forms are
mapped to cocycles (resp. coboundaries). However, note that if $c_k$ is a
cocycle, there are forms $\beta_k$ satisfying (3.1) which are not
closed. All that is required of $d\beta_k$ is that its restriction to the
fibres vanishes.

We would like now to find a two-form on $\sigp$ which is related
to the the two-cocycle (2.13) by the map defined above.
This will require some other preliminaries.
We recall first that there is a map $E$ from $(k+p)$-forms
on $M$ to $k$-forms on $\sigm$ given by
$$
(E\gamma)_\varphi(v_1,\ldots,v_k)=
\int_\Sigma\varphi^*\gamma(v_1,\ldots,v_k,.,\ldots,.)\ .\eqno(3.2)
$$
The same map can be defined with $M$ replaced by $P$ or $G$.
This map commutes with the exterior differential and hence
defines a map of cohomology classes. It is also dual to the evaluation
map in the sense that
$\int_{N_k}E\gamma=\int_{ev(N_k)}\gamma$,
where $N_k$ is a $k$-dimensional submanifold of $\sigm$ and
its evaluation is a $(k+p)$-dimensional submanifold of $M$.
Due to this property, $E$ maps integral cocycles to integral cocycles.

Let us recall also the definition of the transgression in the
bundle $P$. Consider a closed $k$-form $\beta_G$ in the fiber $G$
which is the restriction of a  $k$-form $\beta$ in $P$
such that $d\beta=\pi^*\gamma$ for some
$(k+1)$ form $\gamma$ in $M$. Then one says that $[\gamma]\in
H^{k+1}(M)$ is the transgression of $[\beta_G]\in H^k(G)$.
The classic example of transgression is provided by the Chern--Simons
form $\omega^0_{p+2}(\alpha)$. Since the restriction of a connection
form $\alpha$ to a fiber coincides with the left-invariant
Maurer--Cartan form $g^{-1}dg$, the restriction of
$\omega^0_{p+2}(\alpha)$ to a fiber is
$$
\sigma=\omega^0_{p+2}(g^{-1}dg)=k_p a_p \tr(g^{-1}dg)^{p+2}\
,\eqno(3.3)
$$
where $a_p=(-1)^{{p+1}\over2}
\left({{p+3}\over2}\right)\Gamma({{p+3}\over2})^2/\Gamma(p+3)$.
{}From (2.4) (with $\alpha$ and $\phi$ replacing $A$ and $F$)
there follows that $[\sigma]\in H^{p+2}(G)$ transgresses
to $[c]\in H^{p+3}(M)$.

Our construction will be based on the image of this argument under the map
$E$. The only complication is that the restriction of
$E\omega^0_{p+2}(\alpha)$ to a fiber of $\sigp$ is in general not equal to
$E\sigma$, since the differentials of a general map $\varphi$ are not vertical.
For this reason, it will be convenient to choose a basepoint $\varphi_0$
in $\sigp$ which maps $\Sigma$ to the fiber of $P$ over the basepoint
$x_0$ of $M$. Thus the composition of $\varphi_0$ with the projection
$\pi:P\rightarrow M$ is the constant map $x_0$.
We will call the fiber through $\varphi_0$ the ``typical fiber'', and
identify it with the group $\sigg$.
The restriction of $E\omega^0_{p+2}(\alpha)$ to the typical fiber
is equal to $E\sigma$, because the differentials
of maps in the typical fiber are vertical in $P$.

A direct calculation shows that under the map defined by (3.1) the two-form
$E\omega^0_{p+2}$ is mapped to the cocycle $-\hat c_2$.
In fact, using that on a fundamental vectorfield
$\alpha(\tilde X)=X$ and $\phi(\tilde X,.)=0$, one finds
$$
\eqalign{
(E\omega^0_{p+2}(\alpha))_\varphi(\tilde X,\tilde Y)=
&-(p+2)k_p a_p
\int_\Sigma\varphi^*\tr\left(XY\alpha^p-X\alpha Y\alpha^{p-1}+\cdots
-X\alpha^pY\right)\cr
=&-\hat c_2(\varphi^*\alpha;X,Y)\ .\cr}\eqno(3.4)
$$
It was shown in (2.10) that $\hat c_2$ is cohomologous to $c_2$.
We can now find a two-form $\psi$ which is related to the cocycle
$c_2$.
Consider the one-form $\Upsilon$ on $\sigp$ defined by
$\Upsilon(v)=\int_\Sigma \tr\alpha(v)\Phi_p(\alpha)$.
When evaluated on a fundamental vectorfield, this form is seen to
correspond to the cochain $\Phi$. Therefore,
the form
$$
\psi=-(E\omega^0_{p+2}(\alpha)+d\Upsilon) \eqno(3.5)
$$
is related to the cocycle $c_2$ as in (3.1).
We emphasize at this point that the form $\psi$ is neither closed
nor integral. However, its restriction to the fibers is closed and
integral. It is determined by $c_2$ only up to a form whose contraction
with a vertical vector field is zero.
In particular, it is defined up to a basic form.

In all of the above the connection $\alpha$ in $P$ was kept fixed.
We can now prove that the cohomology class of the extension does not
depend upon the choice of connection.
Let $\psi$ and $\psi'$ be constructed as above starting from two
different connections $\alpha$ and $\alpha'$.
The difference $\psi'-\psi$ is equal to
$E(\omega^0_{p+2}(\alpha')-\omega^0_{p+2}(\alpha))$ plus an exact form.
{}From equation (2.4) and the definition of the Weil homomorphism
follows that
$\omega^0_{p+2}(\alpha')-\omega^0_{p+2}(\alpha)$
is the sum of a basic and a closed form.
Using the properties of $E$, also
$E(\omega^0_{p+2}(\alpha')-\omega^0_{p+2}(\alpha))$
is the sum of a basic and a closed form.
Therefore the restriction of $\psi'-\psi$ to a fiber of $\sigp$
is closed. In fact, it has to be exact. This is because the
restriction of $\omega^0_{p+2}(\alpha')-\omega^0_{p+2}(\alpha)$
to any fiber of $P$ is zero, and therefore the restriction of
$E(\omega^0_{p+2}(\alpha')-\omega^0_{p+2}(\alpha))$ to the typical
fiber defined above is also zero. Forms in $\sigg$ obtained by
restriction of a form on $\sigp$ to different fibers have to
be cohomologous. Therefore the restriction of $\psi'-\psi$ to any
fiber of $\sigp$ has to be exact.
This proves that the cocycles $c_2(\varphi^*\alpha)$
and $c_2(\varphi^*\alpha')$ are cohomologous.

\beginsection{4. EXTENSIONS OF $\sigg$}

We are now going to assume that $\sigg$ is simply connected.
This is the case if $\Sigma$ is the unit sphere $S^p$ and $G$ is
connected, simply connected and has $\pi_{p+1}(G)=0$
(the most important example being $G=SU(N)$ with $N\geq{p+3\over2}$).
Then, there is a one-to-one correspondence between extensions of
the Lie algebra $\Sigma\lieg$ and extensions of the Lie group $\sigg$.
The extensions of $\Sigma\lieg$ by $Maps(\ca,i\br)$ give rise to
topologically nontrivial extensions of $\sigg$ by the abelian group
$Maps(\ca,S^1)$ which have been described in [7].
We are now going to describe the extension $\siggh_\alpha$
corresponding to the Lie algebra $\widehat{\Sigma\lieg}_\alpha$.

We begin by defining $N=\Sigma\times [0,1]$ and fix a basepoint
in each connected component of $\sigg$.
We assume that the basepoints have been chosen in such a
way that the product of two basepoints is another basepoint.
The basepoint in the connected component of the identity has to be the
identity map.
One can extend every map $g:\Sigma\rightarrow G$ to a map
$\hat g:N\rightarrow G$, such that $g(\sigma,1)=g(\sigma)$ and $g(\sigma,0)$
is the basepoint in the connected component containing $g$.
One can also think of $\hat g$ as a path in $\sigg$ beginning at
the basepoint and ending at $g$.
Consider pairs $(\hat g,\lambda)$, where $\hat g\in NG\equiv Maps(N,G)$ and
$\lambda\in Maps(\sigp,S^1)$. We define an equivalence relation
on these pairs: $(\hat g,\lambda)\sim(\hat g',\lambda')$ if
the two paths have the same endpoints $\hat g(1)=\hat g'(1)=g$
and $\lambda'(\varphi)=\lambda(\varphi)e^{2\pi i\int_S \psi}$,
where $S=S[\varphi,\hat g,\hat g']$ is a two-dimensional surface in
$\sigp$ bounded by the paths
$\varphi\hat g$ and $\varphi\hat g'$, both originating at $\varphi$
and ending at $\varphi g$.
Note that one can choose $S$ to lie entirely in the fiber through
$\varphi$ and therefore the integral is not affected by the
arbitrariness in $\psi$.
The group $\siggh_\alpha$ consists of these equivalence classes
of pairs, with the multiplication:
$$
[(\hat g_1,\lambda_1)][(\hat g_2,\lambda_2)]=
[(\hat g_1\hat g_2,\lambda_1(\hat g_1\cdot\lambda_2)
e^{2\pi i\theta_2(.,\hat g_1,\hat g_2)})]\ ,\eqno(4.1)
$$
where $\theta_2$ is a suitable functional of $\varphi$, $\hat g_1$ and
$\hat g_2$ and the left action of $NG$
on $Maps(\sigp,S^1)$ is defined by
$(\hat g\cdot\lambda)(\varphi)=\lambda(\varphi g)$.
For associativity, $\theta_2$ has to be a two-cocycle:
$$
\theta_2(\varphi\hat g_1,\hat g_2,\hat g_3)-
\theta_2(\varphi,\hat g_1\hat g_2,\hat g_3)+
\theta_2(\varphi,\hat g_1,\hat g_2\hat g_3)-
\theta_2(\varphi,\hat g_1,\hat g_2)\equiv 0\quad{\rm mod}\ \bz\ .\eqno(4.2)
$$
In order to reproduce the infinitesimal cocycles (2.4),
the phase has to be chosen as
$$
\theta_2(\varphi,\hat g_1,\hat g_2)=
\int_K \psi\ \eqno(4.3)
$$
where $K$ is the two-dimensional simplex in $\sigp$ with vertices
in $\varphi$, $\varphi g_1$, $\varphi g_1 g_2$ and bounded by the
curves $\varphi \hat g_1$, $\varphi\hat g_1\hat g_2$ and
$\varphi g_1\hat g_2$ (with the obvious notation $(\hat g_1
\hat g_2)(t)=\hat g_1(t)\hat g_2(t)$).
The associativity is then automatically satisfied. In fact
if we call $\tau$ a locally defined one-form in the fiber through
$\varphi$ such that $d\tau$ coincides with the restriction of $\psi$
and define $\theta(\varphi,\hat g)$ to be the line integral
$\int_{\varphi\hat g} \tau$ along the path $\varphi \hat g$,
then
$$
\theta_2(\varphi,\hat g_1,\hat g_2)=
\theta(\varphi \hat g_1,\hat g_2)-
\theta(\varphi,\hat g_1\hat g_2)+
\theta(\varphi,\hat g_1)\ .\eqno(4.4)
$$
The proof that the multiplication (4.1) is independent of the
representatives on the l.h.s. boils down to showing that
$$
\int_{S[\varphi,\hat g_1\hat g_2,\hat g_1'\hat g_2]}\psi-
\int_{S[\varphi,\hat g_1,\hat g_1']}\psi-
\theta_2(\varphi,\hat g_1',\hat g_2)+
\theta_2(\varphi,\hat g_1,\hat g_2)=0\quad{\rm  mod}\ \bz \eqno(4.5)
$$
and
$$
\int_{S[\varphi g_1,\hat g_2,\hat g_2']}\psi-
\int_{S[\varphi,\hat g_1\hat g_2,\hat g_1\hat g_2']}\psi-
\theta_2(\varphi,\hat g_1,\hat g_2)+
\theta_2(\varphi,\hat g_1,\hat g_2')=0\quad{\rm  mod}\ \bz \ .
\eqno(4.6)
$$
These relations are indeed true for $\theta_2$ defined as in (4.3),
since their l.h.s. are the integrals of $\psi$ on closed
two-dimensional submanifolds in a fiber of $\sigp$, and the
restriction of $\psi$ to the fiber is closed and integral.
Note also that $\theta_2$ is not affected by the arbitrariness
in $\psi$ which was discussed in the previous section.

The fiber bundles $\siggh_{\alpha}\to \sigg$ may or may not be trivial;
this depends on the choice of $G$ and $\Sigma.$ For example, when $G=SO(N)$
or any closed subgroup of $SO(N)$ and $p=4n-1$ the bundles become trivial.
The reason is simply that the
characteristic classes tr$\phi^{2n+1}$ vanish
identically when the curvature $\phi$ takes
values in the Lie algebra of antisymmetric real matrices. On the other hand,
when $G=SU(N)$ and $p\leq 2N-3$ then the characteristic classes
tr$\phi^{(p+3)/2}$ are non trivial. In these cases the Chern-Simons form
$\omega^0_{p+2}(\alpha)$, when restricted to a fiber of $P$, coincides
with the form $\sigma$ defined in (3.3) (the WZWN anomaly);
this in turn implies that the form $\psi$ used in the definition of the
extension is topologically nontrivial.

\beginsection{5. THE CIRCLE BUNDLE $\sigpt_{(\alpha,H)}$}

In the introduction we have defined a ``$p$-brane structure''
in $P$ to be a circle bundle over $\sigp$
together with a faithful right action of $\siggh_\alpha$
covering the action of $\sigg$ on $\sigp$.
We will now give a construction of a ``$p$-brane structure''
based on the assumption that the characteristic class
(1.2) vanishes. It seems plausible that $[c]=0$ is also a
necessary condition for the existence of such a structure,
but we shall not investigate this question here.

Assume there exists a basic $(p+2)$-form $H$ on $P$ such that $c=dH$. Then
$$
\Omega=\omega^0_{p+2}(\alpha)-H \eqno(5.1)
$$
is a closed $(p+2)$-form on $P$. One can choose $H$ so that $\Omega$
is also integral [8].
{}From (3.5) and the properties of $E$ we find that the two-form
$\Theta$ in $\sigp$ defined by
$$
\Theta=-(E\Omega+d\Upsilon)=\psi+EH \eqno(5.2)
$$
is closed, integral and is related to $c_2$ by the map (3.1).
If $\sigp$ is simply connected, $\Theta$ can be used to construct a
circle bundle with connection $\sigpt_{(\alpha,H)}$ over $\sigp$.

Again we denote $N=\Sigma\times [0,1]$ and choose a basepoint in each
connected component of $\sigp$. We assume that each connected
component of $\sigp$ is simply connected.
We extend the maps $\varphi:\Sigma\rightarrow P$ to maps
$\hat\varphi:N\rightarrow P$,
which can be regarded as paths in $\sigp$ from the basepoint
$\varphi_0$ to $\varphi$.
The total space of the bundle $\sigpt_{(\alpha,H)}$
is $(Maps(N,P)\times S^1)/\sim$,
where $\sim$ is the equivalence relation given by
$$
(\hat\varphi,\lambda)\sim(\hat\varphi',\lambda e^{2\pi
i\int_S\Omega})\eqno(5.3)
$$
where $\hat\varphi(1)=\hat\varphi'(1)=\varphi$ and
$S=S[\hat\varphi,\hat\varphi']$ is a
two-dimensional submanifold of $\sigp$ bounded by the paths
$\hat\varphi$ and $\hat\varphi'$.
Such a manifold always exists since we have assumed that $\sigp$ is
simply connected and therefore any two homotopies with the same
endpoints can be deformed into each other.
We denote points of $\sigpt_{(\alpha,H)}$ by equivalence classes
of pairs $[(\hat\varphi,\lambda)]$.

We define a right action of $\siggh_\alpha$ on $\sigpt_{(\alpha,H)}$ by
$$
[(\hat\varphi,\lambda)]\cdot[(\hat g,\mu)]=
[(\hat\varphi\cdot \hat g,(\lambda\cdot \hat g)\mu
e^{2\pi i\chi(\hat\varphi,\hat g)})]\ .\eqno(5.4)
$$
The phase factor is defined by
$$
\chi(\hat\varphi,\hat g)=\int_K\Theta\ ,\eqno(5.5)
$$
where $K=K[\hat\varphi,\hat g]$ is the two-simplex in $\sigp$
with vertices in $\varphi_0$, $\varphi$, $\varphi g$
and bounded by the paths
$\hat\varphi$, $\varphi\hat g$ and $\hat\varphi\hat g$. Note that
unlike in section 4, the phase is now defined as an integral of
$\Theta$ rather than $\psi$. This is because the surface $K$ does not
lie in a fiber of $\sigp$ and $\psi$ was closed and integral only when
restricted to a fiber.

To prove that this action is independent of the representatives on the
l.h.s. one has to show that
$$
\chi(\hat\varphi',\hat g)-\chi(\hat\varphi,\hat g)-
\int_{S[\hat\varphi\hat g,\hat\varphi'\hat g]}\Theta
+\int_{S[\hat\varphi,\hat\varphi']}\Theta
=\,0 \quad{\rm{mod}}\ \bz\eqno(5.6)
$$
and
$$
\chi(\hat\varphi,\hat g')-\chi(\hat\varphi,\hat g)-
\int_{S[\hat\varphi\hat g,\hat\varphi\hat g']}\Theta
+\int_{S[\varphi\hat g,\varphi\hat g']}\Theta
=\,0 \quad{\rm{mod}}\ \bz\ .\eqno(5.7)
$$
These relations are indeed true, since their left hand sides are the
integrals of $\Theta$ on closed two-dimensional submanifolds of
$\sigp$.

Finally we have to show that this action is compatible with the group
multiplication given in (4.1). Acting on $[(\hat\varphi,\mu)]$
with $[(\hat g_1,\lambda_1)]$ and then with $[(\hat g_2,\lambda_2)]$,
and comparing with the action of the product of these group elements,
we are led to the condition that the coboundary of $\chi$ has to be
$\theta_2$:
$$
\chi(\hat\varphi\hat g_1,\hat g_2)-
\chi(\hat\varphi,\hat g_1\hat g_2)-
\chi(\hat\varphi,\hat g_1)=
\theta_2(\varphi,\hat g_1,\hat g_2)\quad{\rm mod}\ \bz\ .\eqno(5.8)
$$
This condition is again true, as one can easily verify by considering
the tetrahedron in $\sigp$ with vertices in $\varphi_0$, $\varphi$,
$\varphi g_1$, $\varphi g_1 g_2$ and with edges $\hat\varphi$,
$\hat\varphi\hat g_1$, $\hat\varphi\hat g_1\hat g_2$,
$\varphi\hat g_1$, $\varphi\hat g_1\hat g_2$, $\varphi g_1 \hat g_2$.
The three faces touching $\varphi_0$ give the three terms on the
l.h.s.. The remaining face lies entirely in the fiber through
$\varphi$ and therefore the
integral of $\Theta$ on that face is equal to the integral of $\psi$,
thus reproducing (4.3).

\beginsection{6. A MODEL OF BOSONIC $p$-BRANES}

Fix a riemannian metric $g$ on $M$ and an invariant positive definite
bilinear form $\langle.,.\rangle$ in the Lie algebra $\lieg$.
For any principal connection in $P$ one can construct a unique
riemannian metric on $P$ such that the following holds:
i) the horizontal and vertical spaces are orthogonal, ii) the
inner product of horizontal vectors is equal to the inner product of
their projections to $M$ and iii) the inner product of vertical
vectors is equal to the inner product of the corresponding
Lie algebra elements.

The manifold $\Sigma$ is endowed with a riemannian metric $\gamma$.
The time evolution of the
$p$-brane is given by a map, still denoted $\varphi$, from
$\Sigma\times I$ to $P$, where $I$ is a time interval.
The action is the sum of a kinetic and a topological part:
$S=S_{\rm{kin}}+S_{\rm{top}}$.
The kinetic term is
$$
S_{\rm{kin}}={1\over2}
\int_{\Sigma\times I}d^p\sigma dt \sqrt{-\rm{det}\gamma}
\gamma^{ij}(\partial_i\varphi,\partial_j\varphi)_P\ ,\eqno(6.1)
$$
where $(.,.)_P$ is the riemannian inner product defined above and the
riemannian metric $\gamma$ of $\Sigma$ has been extended to a product
Lorentzian metric on $\Sigma\times I.$
In order to define the topological term we assume that space-time
is compact and without boundary. This can be done by compactifying
$I$ to $S^1$. Then
$$
S_{\rm{top}}=\int_{N\times S^1}\hat\varphi^\ast\Omega\eqno(6.2)
$$
where $\Omega$ has to be a closed integral $(p+2)$-form on $P$.
At each time the field $\varphi$ is extended to a map $N\to P$ as before.
We choose $\Omega$ to be as in (5.1).

To make contact with earlier work, we compute the form of the action
with respect to a local trivialization of $P$.
If $(x,h)$ are local coordinates on $P$, the connection form
is given by
$$
\alpha(x,h)=h^{-1}dh+h^{-1}A(x)h\ ,\eqno(6.3)
$$
where $A$, a locally defined form on $M$, is the Yang--Mills potential.
Locally, the map $\varphi$ can be represented by a pair of maps
$x:\Sigma\rightarrow M$ and $h:\Sigma\rightarrow G$.
The differential of $\varphi$ can be decomposed into vertical
and horizontal parts:
$d\varphi=(dx,dh)=(0,dh+h\cdot x^\ast A)+(dx,-h\cdot x^\ast A)$.
Inserting into (6.1)
$$
\eqalign{
S_{\rm{kin}}={1\over2}\int_{\Sigma\times I}d^p\sigma dt &
\sqrt{-\rm{det}\gamma}\gamma^{ij}\bigl[g(\partial_i x, \partial_j x)\cr
&+ \langle h^{-1}\partial_i h+\partial_i x^\mu A_\mu,
h^{-1}\partial_j h+\partial_j x^\mu A_\mu\rangle \bigr].}\eqno(6.4)
$$
For the topological term we observe that using (6.3)
$$
\omega^0_{p+2}(\alpha)=
\omega^0_{p+2}(A)+dC(A,dhh^{-1})+\omega^0_{p+2}(dhh^{-1})\ ,\eqno(6.5)
$$
where the $p+1$ form $C$ is a differential polynomial in the indicated
arguments.
Furthermore, we can define locally a $(p+1)$-form $B$ on $M$
such that
$$
H=\omega^0_{p+2}(A)+dB\ .\eqno(6.6)
$$
Then
$$
\Omega=\omega^0_{p+2}(dhh^{-1})+dC-dB\ .\eqno(6.7)
$$
This is the topological term given in [3] (in comparing with [3]
one has to take into account that here we are gauging the left
action of $G$ on itself whereas in [3] the right action was gauged,
contrary to what is stated there).

Strictly speaking in the case $p>1$ the only invariance of the action
is the finite dimensional group $G$ acting on $\varphi$ by right
multiplication. This is because the connection $\alpha$ is not a
dynamical variable and therefore should be treated as a fixed
background. However, if one transforms also $\alpha$, the action is
also invariant under the group $\cg=\aut P$ acting on $\varphi$ by
composition and on $\alpha$ and $H$ by pullback (in particular, since $H$
is basic, it is invariant).

In the Schr\"odinger picture the wave functions are
sections of a complex line bundle $\cal L$ over $\sigp$.
The choice of the (equivalence class of the) line bundle is dictated
by the action principle. In the present case the line bundle is obtained
as the associated bundle to the circle bundle $\sigpt_{(\alpha,H)}$ through
the natural representation of $S^1$ in $\bf C.$ This can be checked by
comparing the anomalies of the Poisson brackets of Noether charges
associated to the symmetry group of right $G$ multiplications on $P$
and the commutator anomalies (2.11).
It was found in [5] that in the case of a trivial bundle $P=M\times G$,
the Poisson bracket algebra of the Noether charges associated to the right
action of $G$ on $P$ has an extension given by the cocycle $c_2(h^{-1}dh)$.
On the other hand, we recall from section 3 that the cohomology class of
the extension $\widehat{\Sigma\lieg}_\alpha$ is independent of the
connection $\alpha$.
Thus if $P$ is a trivial bundle one can choose without loss of
generality the flat connection for which $\alpha$ is equal to the
left-invariant Maurer--Cartan form. This means that if one wants to
lift the action of $\sigg$ on $\sigp$ to the quantization bundle,
the group has to be extended to $\siggh_{h^{-1}dh}$.
Thus the quantization bundle is $\sigpt_{(h^{-1}dh,H)}$.
Essentially the same local computations apply,
using local trivializations, in the general case.

Up to this point we have considered the construction of the
circle bundle $\sigpt_{(\alpha,H)}$ for fixed $\alpha$ and $H$.
Let ${\cal M}$ be the set of all pairs $(\alpha,H)$ such that
$\omega^0_{p+2}(\alpha)-H$ is closed and integral.
Now consider the union of all bundles $\sigpt_{(\alpha,H)}$;
since the construction of these bundles depended smoothly on
the data $(\alpha,H)$, we obtain a bundle $\sigpt$
over ${\cal M}$ with fibers $\sigpt_{(\alpha,H)}$.
We can represent points of $\sigpt$
as quadruples $(\hat\varphi,\lambda,\alpha,H)$, with the equivalence
relation $(\hat\varphi,\lambda,\alpha,H)\sim
(\hat\varphi',\lambda e^{2\pi i\int_S\Omega},\alpha,H)$,
where $S$ is as in (5.3).
Note that $\sigpt$ is also a circle bundle over $\sigp\times{\cal M}$.

The gauge group $\cg=\aut P$ acts on $\sigp$ by composition:
for $u\in\cg$, $(u\cdot\varphi)(\sigma)=u(\varphi(\sigma))$.
This definition can be applied also to maps from $N$ to $P$.
The automorphisms act also on forms on $P$ by pullback. In particular,
this gives the usual action on connections, and a trivial action
on basic forms such as $H$.

We would like to lift this action of $\cg$ on $\sigp\times{\cal M}$
to an action on $\sigpt$.
The obvious definition of the action of $\cg$ is
$$ u[(\hat\varphi,\lambda,\alpha,H)]=
[(u\cdot\hat\varphi,\lambda,u^{-1*}\alpha,H)].\eqno(6.8)$$
If we apply this to equivalent quadruples we find that since
$\omega^0_{p+2}(u^{-1*}\alpha)=u^{-1*}\omega^0_{p+2}(\alpha)$,
also the transformed quadruples are equivalent.
Therefore the definition given above goes to the quotient and
defines an action of $\cg$ on $\sigpt$.
In this case there is no extension, and at the infinitesimal level
no commutator anomaly. Again this agrees with the result of [4,5]
for a trivial bundle.
Note that if one writes $H$ locally in the form (6.6), the gauge
variations of the two terms on the r.h.s. have to cancel. Using
(2.5), the gauge variation of $B$ has to be
$\delta_X B=-\omega^1_{p+1}(A,X)$, up to an exact form.

\vskip 0.7in
\centerline{\bf REFERENCES}

\item{[1]} T.P. Killingback, Nucl. Phys. \bf B288 \rm (1987) p. 578.
\item{[2]} R. Coquereaux and  K. Pilch,
Commun. Math. Phys. \bf 120 \rm (1989) p. 353.
\item{[3]} J.A. Dixon, M.J. Duff and E. Sezgin,
Phys. Lett. \bf 279 B \rm (1992) p.265.
\item{[4]} E. Bergshoeff, R.Percacci, E. Sezgin, K.S. Stelle
and P.K. Townsend, ``$U(1)$-extended gauge algebras in p-loop space'',
SISSA 157/92/EP, to appear in Nucl. Phys.
\item{[5]} R. Percacci and E. Sezgin, ``Symmetries of p-branes'',
SISSA 182/92/EP.
\item{[6]} J. Mickelsson,  Lett. Math. Phys. {\bf    7} (1983) p. 45;
Commun.   Math.   Phys.   {\bf 97} (1985) p. 361;
L. Faddeev, Phys. Lett. {\bf 145B} (1984) p. 81;
L. Faddeev and S.L. Shatashvili, Theor. Math. Phys. \bf 60 \rm (1984)
p. 770.
\item{[7]}  J. Mickelsson: \it Current Algebras and Groups. \rm Plenum Press,
London and New York (1989); Commun. Math. Phys. \bf 110 \rm (1987) p. 173.
\item{[8]} S.-S. Chern and  J. Simons, Ann. of Math. \bf 99 \rm (1974) p. 48.

\bye